\documentclass[letter,twocolumn]{jpsj2}

\title{Creation of superconducting vortices by angular momentum of light}

\author{Takehito Yokoyama$^1$\thanks{yokoyama@stat.phys.titech.ac.jp}}
\inst{$^1$Department of Physics, Tokyo Institute of Technology, Tokyo 152-8551,
Japan } 

\abst{We investigate a superconducting state irradiated by a laser beam with spin and orbital angular momentum. It is shown that superconducting vortices are created by the laser beam due to heating effect and transfer of angular momentum of light. Possible experiments to verify our prediction are also discussed.}

\begin{document}
\maketitle

Interaction of light with condensed matter systems has offered a new approach to control and study them in an ultrafast manner.\cite{Kirilyuk,Lenk,Kirilyuk2,Walowski}
Various aspects of superconductors under light illumination have been investigated such as response of an Abrikosov vortex\cite{Janko},  superconducting single photon detectors\cite{Maingault,Engel}, excitation of Higgs modes\cite{Barankov,Yuzbashyan,Matsunaga,Krull}, laser created Josephson junctions\cite{Magrini}, manipulation of a vortex\cite{Veshchunov}, probe and control of chirality of Cooper pairs\cite{Xia,Vadimov,Claassen}, and excitation of phonons\cite{Schnyder,Vodolazov}, which are fundamentally and technologically important. Most of the previou works have considered energy transfer of light. One of the consequences is the local increase of electron temperature. This effect is captured by the hot spot model where the temperature varies in space.\cite{Zotova,Vadimov2} As a result, the superconductivity can be locally suppressed or destroyed around the hot spot. On the contrary, in this work, we focus on the angular momentum of light.

Light has spin and orbital angular momentum. The spin angular momentum is given by circular polarization. On the other hand, the orbital angular momentum is given by spatial structure of the light: the phase structure of the light is winding and hence the intensity is zero at the center of the beam (along the propagation axis), analogous to superconducting vortices. This type of light is dubbed optical vortices.\cite{Allen,ONeil} There have been various demonstrations of orbital angular momentum transfer, e.g., to classical particles\cite{He,Friese} or excitons\cite{Shigematsu}. It also has been demonstrated that spin angular momentum of circularly polarized light can be used to reverse magnetization.\cite{Stanciu} 

Since the relevant length scales of superconductors and opical vortices are the same ($\sim 1 \mu$m), it can be expected that orbital angular momentum of light can be imprinted on superconductors.

In this paper, we investigate a superconducting state irradiated by a laser beam with spin and orbital angular momentum. It is shown that superconducting vortices are created by the laser beam due to heating effect and transfer of angular momentum of light. Possible experiments to verify our theory are also discussed.

\begin{figure}[tbp]
\begin{center}
\scalebox{0.8}{
\includegraphics[width=8.50cm,clip]{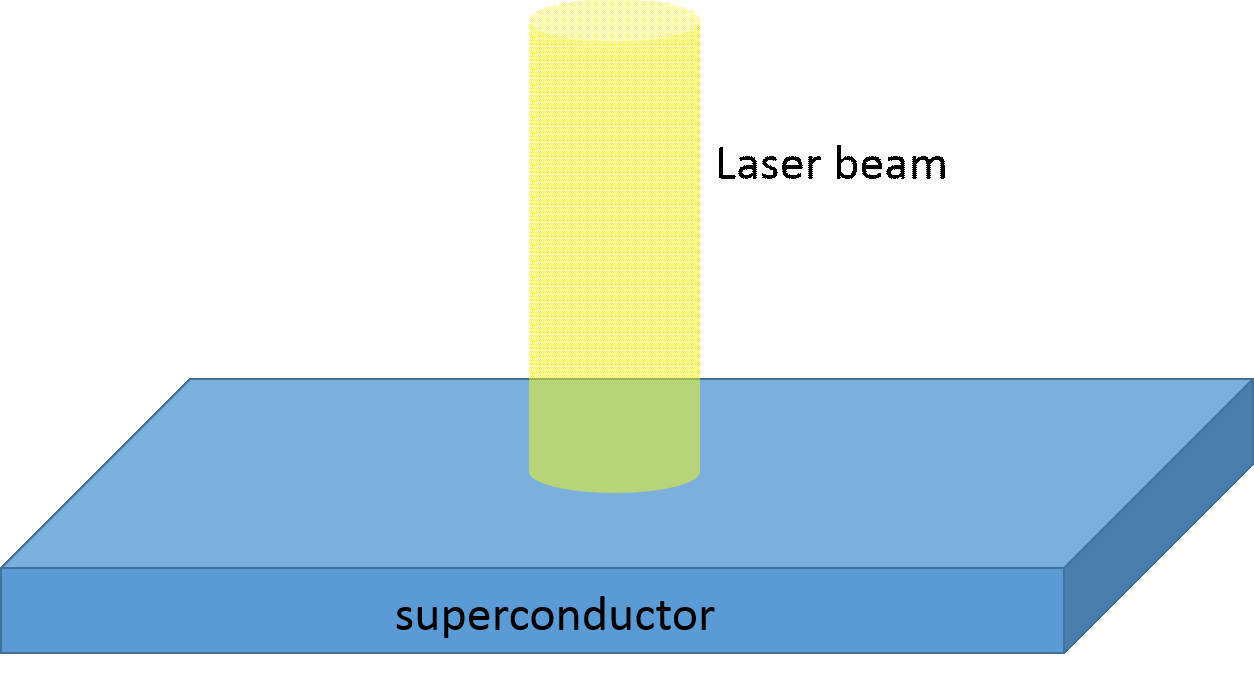}
}
\end{center}
\caption{(Color online) Schematic illustration of a superconductor irradiated by a laser beam.}
\label{fig1}
\end{figure}

We consider a two dimensional superconductor irradiated by a laser beam with spin and orbital angular momentum as shown in Fig. \ref{fig1}. 
We assume that photon is completely absorbed by the superconductor without any reflection and transmission of the light. The dynamics of the superconductor can be described by the Ginzburg-Landau theory.

The Ginzburg-Landau free energy of this system reads
\begin{eqnarray}
F = \int_{}^{} {\left\{ {\alpha \left[ {T({\bf{r}},t) - {T_c}} \right]|\psi {|^2} + \gamma {{\left| {\left( {\nabla  + i\frac{{2e}}{\hbar }{\bf{A}}} \right)\psi } \right|}^2}} \right\}} {d^2}{\bf{r}}.
\end{eqnarray}
$T_c$ is the transition temperature of the superconductor. Due to the heating effect of the laser beam, the temperature $T$ depends on space and time.\cite{Zotova,Vadimov2}
The time dependent Ginzburg-Landau (TDGL) equation is given by $ - \Gamma \frac{{\partial \psi }}{{\partial t}} = \frac{{\delta {\cal F}}}{{\delta {\psi ^*}}}$.\cite{Ivlev}
With the proper normalization\cite{Magrini}, it reduces to
\begin{eqnarray}
\frac{{\partial \psi }}{{\partial t}} = \tau ({\bf{r}},t)\psi  + {(\nabla  + i{\bf{A}})^2}\psi 
\end{eqnarray}
with $\tau ({\bf{r}},t) =(T({\bf{r}},t) - T_c)/(T_\infty - T_c)$ and $T({\bf{r}},t) = {T_\infty } + \Delta T({\bf{r}},t)$. Here, ${T_\infty }$ and $\Delta T({\bf{r}},t)$ denote the temperature far from the laser beam and the variation of the temperature by the laser beam, respectively. The typical time scale of the TDGL equation is ps. The length is normalized by the superconducting coherence length $\xi \sim 1$ $\mu$m.
We consider a laser pulse of the form
${\bf{A}} = \frac{{2e}}{\hbar }\frac{i}{\omega }\xi {\bf{E}}\delta (t)$. Here, $\omega$ is the frequency of the laser beam. 
We assume that the thermalization time due to energy dissipation is much shorter than the other characteristic times and the relaxation time for angular momentum is much longer than the other characteristic times. Thus, energy transfer is neglected but transfer of angular momentum is taken into account in the vector potential. We neglect the ${{\bf{A}}^2}$ term since this term does not transfer angular momentum and hence does not change the results qualitatively. We also assume that the generated vortex is pinned by a defect or impurity.

With the initial condition $\psi ({\bf{r}},0) = {\psi _b}$ (constant), we then obtain
\begin{eqnarray}
\frac{{\partial \psi }}{{\partial t}} = \tau ({\bf{r}},t)\psi  + {\nabla ^2}\psi  + i(\nabla  \cdot {\bf{A}}){\psi _b}.
\end{eqnarray}
We expand the order parameter in a Fourier series as
$\psi  = \sum\limits_n {{e^{in\theta }}{\psi _n}(r)}$.

We use two types of laser beams: left circularly polarized Gaussian and Bessel beams. The Bessel beam carrys orbital angular momentum.
First, consider a left circularly polarized Gaussian beam
\begin{eqnarray}
{\bf{E}} ={E_0}\exp \left( { - \frac{{{r^2}}}{{2{\sigma ^2}}}} \right) \left( {\begin{array}{*{20}{c}}
1\\
i
\end{array}} \right)\end{eqnarray}
with a constant $E_0$ and the radius of the beam $\sigma$, 
and $\Delta T({\bf{r}},t) = {T_b}\exp \left( { - \frac{{{r^2}}}{{{\sigma ^2}}}} \right)\delta (t)$. The variation of the temperature is assumed to be proportional to the intensity of the beam with a proportionality coefficient $T_b$.
Then, the TDGL equation becomes
\begin{eqnarray}
\frac{{\partial {\psi _n}}}{{\partial t}} = \tau ({\bf{r}},t){\psi _n} + \nabla _n^2{\psi _n}, \; n \ne 1,
\end{eqnarray}
\begin{eqnarray}
\frac{{\partial {\psi _1}}}{{\partial t}} = \tau ({\bf{r}},t){\psi _1} + \nabla _1^2{\psi _1} + \frac{{2e}}{\hbar }\frac{{{E_0}\xi }}{{\omega {\sigma ^2}}}r\exp \left( { - \frac{{{r^2}}}{{2{\sigma ^2}}}} \right){\psi _b}\delta (t)
\end{eqnarray}
with $\nabla _n^2 = \frac{{{\partial ^2}}}{{\partial {r^2}}} + \frac{1}{r}\frac{\partial }{{\partial r}} - \frac{{{n^2}}}{{{r^2}}}$.
Therefore, we find ${\psi _n} = 0$ for $ n \ne 0,1$.

Next, consider a left circularly polarized Bessel beam of the form\cite{Andrews}
\begin{eqnarray}
{\bf{E}} ={E_0}{J_m}(kr){e^{im\theta }} \left( {\begin{array}{*{20}{c}}
1\\
i
\end{array}} \right)
\end{eqnarray}
and
$\Delta T({\bf{r}},t) = {T_b}{\left( {{J_m}(kr)} \right)^2}\delta (t)$.
Here, $J_m$ is the Bessel function of the first kind and order $m$. $k$ is the wave number of the beam.

The TDGL equation reads
\begin{eqnarray}
\frac{{\partial {\psi _n}}}{{\partial t}} = \tau ({\bf{r}},t){\psi _n} + \nabla _n^2{\psi _n},n \ne m + 1,\\
\frac{{\partial {\psi _{m + 1}}}}{{\partial t}} = \tau ({\bf{r}},t){\psi _{m + 1}} + \nabla _{m + 1}^2{\psi _{m + 1}} \nonumber \\ + \frac{{2e\xi {E_0}k}}{{\hbar \omega }}{J_{m + 1}}(kr)\delta (t).
\end{eqnarray}
We see ${\psi _n} = 0$ for $n \ne 0,m + 1$.

In the numerical calculation, we set $\delta (t) \to \frac{1}{{2{t_0}}}\Theta ({t_0} - \left| t \right|)$ with ${t_0} = {10^{ - 3}},  {\psi _b} = 1$, ${T_\infty }/{T_c}=0.9,$ and $\frac{{2e\xi {E_0}}}{{\hbar \omega }}=10$ which corresponds to ${E_0}$ = 5kV/m, $\omega$=0.25THz, and $\xi$=1$\mu$m.
We also use $\sigma=1$ for the Gaussian beam and $k=1, m=1$ for the Bessel beam.We have solved the TDGL equations with the boundary condition ${\psi _n}(r \to \infty ,t) = 0$ by the Crank-Nicolson method numerically. In the following, the numerical results are shown at $t=2t_0$.


\begin{figure}[tbp]
\begin{center}
\scalebox{0.8}{
\includegraphics[width=9.50cm,clip]{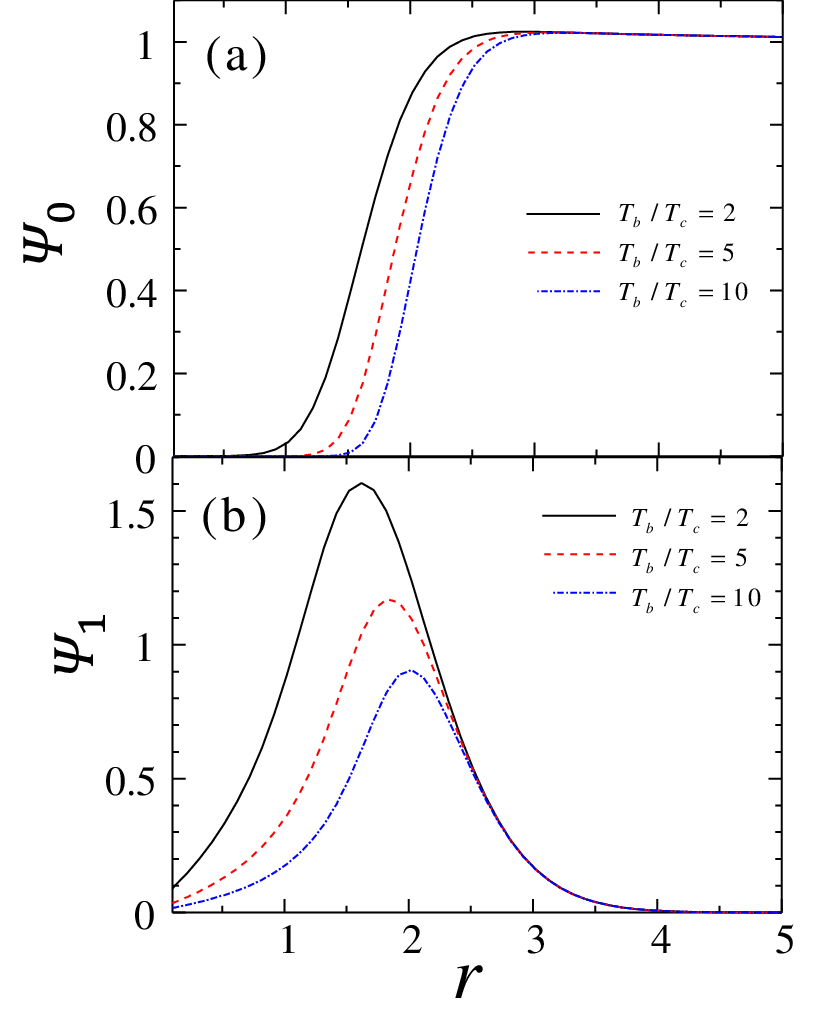}
}
\end{center}
\caption{(Color online)  The order paremeters as a function of $r$ for various $T_b$ under the left circularly polarized Gaussian beam. (a) $\psi_0$. (b) $\psi_1$.}
\label{fig2}
\end{figure}

\begin{figure}[tbp]
\begin{center}
\scalebox{0.8}{
\includegraphics[width=9.50cm,clip]{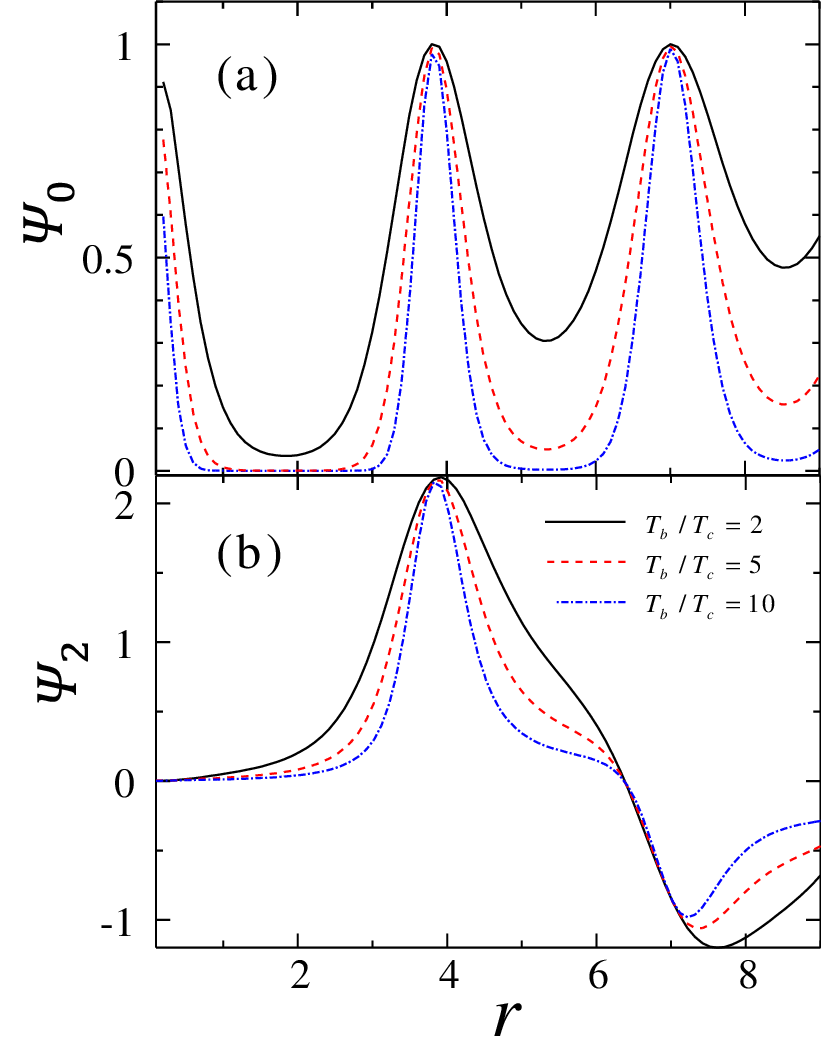}
}
\end{center}
\caption{(Color online)  The order paremeters as a function of $r$ for various $T_b$ under the left circularly polarized Bessel beam. (a) $\psi_0$. (b) $\psi_2$.}
\label{fig3}
\end{figure}

Figure \ref{fig2} shows the order paremeters as a function of $r$ for various $T_b$. It is seen that the bulk component $\psi_0$ is reduced over the laser width ($\sigma=1$) as shown in Fig. \ref{fig2} (a). The order parameter with vorticity one $\psi_1$ is induced by the laser beam as shown in Fig. \ref{fig2} (b). Away from the laser beam, this component goes to zero. The TDGL equation for $\psi_1$ contains the derivative of the electric field. Thus, $\psi_1$ has a peak around the edge of the beam.
These components are reduced with the increase of $T_b$.
Here, we have considered a left circularly polarized beam. For a right circularly polarized beam, the results are identical except for the replacement $\psi_1 \to \psi_{-1}$.

Figure \ref{fig3} depicts the order paremeters as a function of $r$. It is seen that the bulk component is reduced, reflecting the spatial profile of the Bessel function as seen in Fig. \ref{fig3} (a).
Since $J_1(0)=0$, $\psi_0$ is not so reduced at the center of the beam $r=0$. The order parameter with vorticity 2 is induced by the laser beam as shown in Fig. \ref{fig3} (b).
Since $J_1(r)$ is an oscillating function of $r$, these components also oscillate. In particular, $\psi_2$ changes its sign. This is because heating acts on both $\psi_0$ and $\psi_2$ but $\psi_2$ is also affected by $\nabla \cdot {\bf{A}}$ term which changes its sign, as seen from Eq.(9). 
It is remarkable that these components have ring-shaped spatial profiles.
The irradiation of optical vortices makes it possible to create vortices with vorticity more than one, so-called giant vortices\cite{Schweigert,Bruyndoncx}.
Vortex with vorticity more than one is usually unstable and split into vortices with vorticity one.

In general, for light with spin and orbital angular momenta $s$ and $l$, the superconductor acquires the angular momentum $s+l$.
The vortex created by a laser beam is characterized by vorticity or winding number. Thus, once created, it is topologically stable and does not vanish spontaneously.
For time evolution of the order paremeters, see Supplemental Materials.\cite{SM}

As for the generation of the order parameters with vorticity, two mechanisms work: (i) heating breaks superconductivity locally; (ii) angular momentum of light is transferred to the superconductor. These mechanisms play a role analogous to local breakdown of superconductivity by magnetic field and flux quantum trapped in the vortex core.

\begin{figure}[tbp]
\begin{center}
\scalebox{0.8}{
\includegraphics[width=9.50cm,clip]{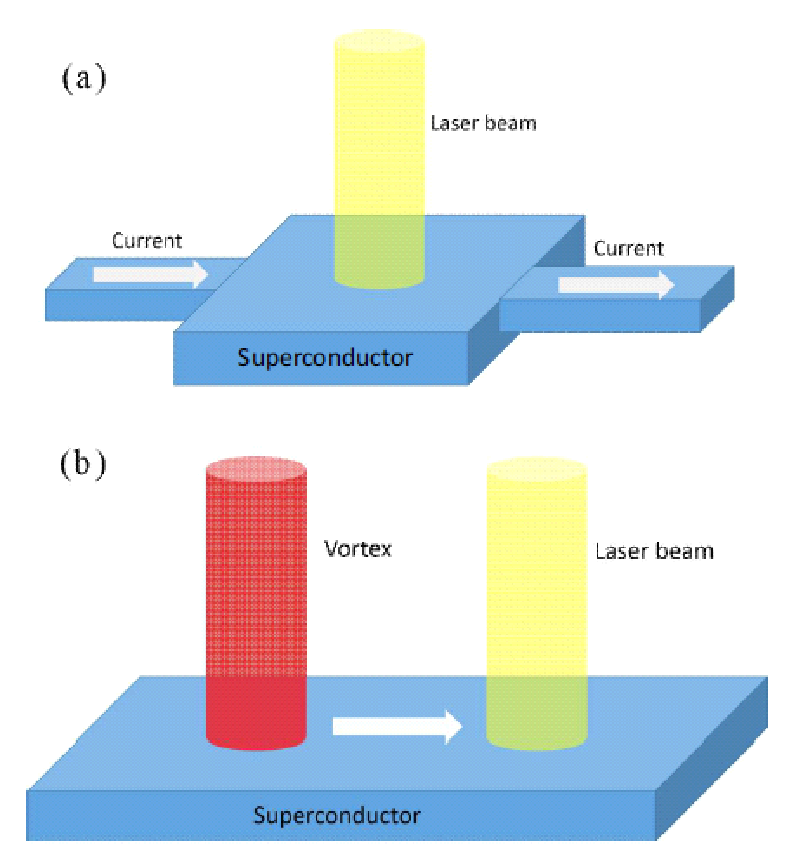}
}
\end{center}
\caption{ (Color online) Experimental setups to verify our theory. (a) Laser created superconducting quantum interference device can be used to measure vorticity or magnetic flux at the vortex transferred from the light. (b) Motions of vortices around the laser spot reflect the angular momentum of light.}
\label{fig4}
\end{figure}

Now, let us discuss possible experiments to verify our prediction.  Since a laser beam can create a superconducting vortex, laser illumination around the center of the superconductor of finite size can create superconducting quantum interference device as shown in Fig. \ref{fig4} (a). This way, one can measure vorticity or magnetic flux at the vortex transferred from the light. 
In the mixed phase under the temperature gradient, the vortices experience the force proportional to the temperature gradient.\cite{Veshchunov} When the vorticities of the vortices have the same (opposite) sign, the interaction between them is repulsive (attractive). Thus, the total force acting on the vortices and hence their velocities approaching the laser spot depend on the angular momentum of light, as shown in Fig. \ref{fig4} (b), which can be used to check our theory.
In particular, vortices with opposite vorticities, i.e., vortices and antivortices, can be pair annihilated. Therefore, it is also possible to eliminate vortices by laser beam, similar to laser-induced demagnetization in magnets\cite{Kirilyuk,Beaurepaire}. The electric states at the vortex core also reflect vorticity which can be measured by scanning tunneling microscope.

Our results are also applicable to type I superconductors where vortices cannot be generated by magnetic field.
Votices in chiral p-wave superconductors can host Majorana fermions.\cite{Ivanov} When our results are applied to chiral p-wave superconductors, Majorana fermions can be controlled by a laser beam.

We have neglected the nonlinear terms with respect to the order parameter, which lead to couplings between different components of the order parameter ($\psi_n$). Thus, other components than those considered in this paper are also generated in the presence of the nonlinear terms. However, these components are less dominant. 

Finally, let us comment on experimental feasibility. One can use Al, a conventional superconductor with coherence length about 1 micron, as a superconductor. As a substrate, one can use sapphire which has a thermal conductivity much larger than that of the superconductor to minimize the hot spot size\cite{Magrini}. To avoid quasiparticle excitations, one should use subterahertz light. THz laser has been realized as shown in Refs. \cite{Matsunaga,Kim} .
Optical vortex can be generated by several methods such as a spiral phase plate, holograms, metasurfaces or a spatial light modulator\cite{Wang,Shen,Pachava,Forbes}.
Therefore, the proposed system can be prepared with current technology.

In summary, we have investigated a superconducting state irradiated by a laser beam with spin and orbital angular momentum. We have shown that superconducting vortices are created by the laser beam due to heating effect and transfer of angular momentum of light. Experimental setups to verify our theory are also discussed.

The author thanks N. Yoshikawa, S. Okuma, S. Kaneko, and  K. Ienaga for discussions.
This work was supported by JSPS KAKENHI Grant Number JP30578216 and the JSPS-EPSRC Core-to-Core program "Oxide Superspin".

\end{document}